# NMR Studies on the Superconducting Symmetry of Iron Pnictide Systems

Y. Kobayashi,[1,2] A. Kawabata,[1] S. C. Lee,[1] T. Moyoshi,[1] and M. Sato[1,2],*

[1]*Department of Physics, Division of Material Science, Nagoya University, Furo-cho, Chikusa-ku, Nagoya 464-8602*
[2]*JST, TRIP, Nagoya University, Furo-cho, Chikusa-ku, Nagoya 464-8602*

NMR longitudinal relaxation rates $1/T_1$ and Knight shifts $K$ have been measured for superconducting samples of LaFe$_{1-y}$Co$_y$AsO$_{1-x}$F$_x$ with $y=0.0$ and $0.0075$ and for a nonsuperconducting metallic sample with $y=0.1$, where the $x$ values are always fixed at 0.11. The temperature ($T$) dependence the relaxation rates $1/T_1$ of the superconducting samples has been found to be markedly different from the behavior $1/T_1 \propto T^{2.5-3.0}$ reported by many groups in the entire $T$ range measured (from the temperature immediately below the superconducting transition temperature $T_c$ down to $(0.1\sim0.2)T_c$). The nonexistence of the coherence peak has also been found. Based on the results of the measurements and other kinds of existing data, arguments are presented on the superconducting symmetry of the Fe pnictide systems, where the several points which cannot be easily understood by existing theories, are pointed out. Results of the measurements on the nonsuperconducting metallic samples are also presented.

The superconductivity found in LaFeAsO$_{1-x}$F$_x$[1] has attracted much interest from the view points of both pure and applied sciences. Because the superconductivity is realized in the FeAs conducting layers formed of corner-sharing FeAs$_4$ tetrahedra, where the 3$d$ electrons are acting a main role, many groups have studied the spin fluctuation mechanism as a possible candidate and proposed the s$_\pm$-symmetry of the superconducting order parameter $\Delta$,[2,3] which is characterized by opposite signs of the $\Delta$ values on the disconnected Fermi surfaces around $\Gamma$ and M points.

Various results are also being accumulated for the arguments on the symmetry and mechanism. For examples, the macroscopic properties such as the transport behaviors,[4-7] impurity-doping effect on the superconducting transition temperature $T_c$,[4-12] magnetic penetration depth $\lambda$,[13] the Fe isotope effect on $T_c$[14] and so on have been studied. Microscopic quantities have also been measured by NMR,[15-20] neutron magnetic scattering[21-24] and angle-resolved photoemission spectroscopy (ARPES).[25,26] Among various reports made from these experimental studies, we just describe here following tentative results relevant to arguments of the present paper. (i) The suppression of $T_c$ by Co doping is much weaker than that expected for superconductors with node(s) of $\Delta$. As we have stressed,[4-6] the insensitiveness of $T_c$ to Co doping seems not to simply be understood even for superconductors which do not have node on their Fermi surfaces but have order parameters $\Delta$ with opposite signs on disconnected Fermi surfaces.[4-7] (ii) A simple relation $1/T_1 \propto T^{2.5-3.0}$ was observed by many groups between the NMR longitudinal relaxation rate $1/T_1$ and temperature $T$. Although the relation $1/T_1 \propto T^3$ is well known for superconductors with line nodes, it is basically expected just in the low temperature region. In contrast, the $1/T_1 \propto T^{2.5-3.0}$ relation observed for Fe pnictides holds in the almost entire $T$ region experimentally studied (from the temperature immediately below $T_c$ down to $(0.1\sim0.2)T_c$), and therefore the relation $1/T_1 \propto T^{2.5-3.0}$ may be a distinct phenomenon from that of superconductors with line nodes. The so-called coherence peak has not been observed. For proper understanding of the superconductivity, it is important to explain many results consistently. At this moment, it is still unsatisfactory.

In the present paper, we reports results of our NMR studies, mainly on the relaxation rate $1/T_1$ of superconducting samples of LaFe$_{1-y}$Co$_y$AsO$_{1-x}$F$_x$ ($y=0.0$ and $0.0075$) and examine if the s$_\pm$-symmetry can explain the experimental results. We also report NMR results of a nonsuperconducting metallic sample of LaFe$_{1-y}$Co$_y$AsO$_{1-x}$F$_x$ ($y=0.1$) (One can find a brief description on the phase diagram of LaFe$_{1-y}$Co$_y$AsO$_{1-x}$F$_x$, later.).

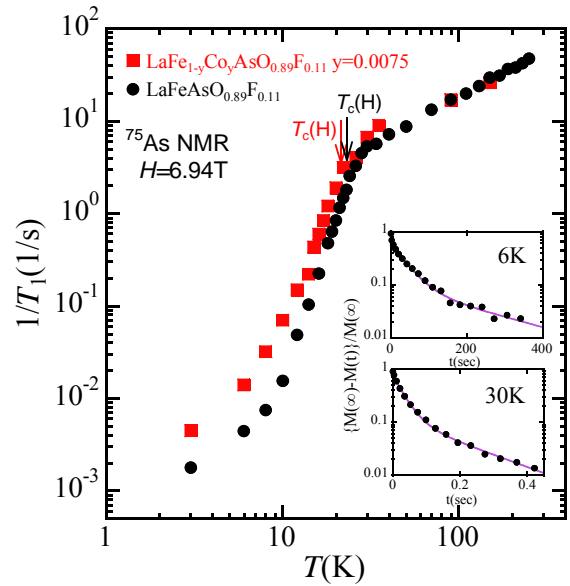

Fig. 1   $^{75}$As NMR longitudinal relaxation rates $1/T_1$ are plotted against $T$ for LaFe$_{1-y}$Co$_y$AsO$_{1-x}$F$_x$ ($y=0.0$ and $0.0075$) with the logarithmic scales, where the $T_c$ values in the magnetic field are shown by the arrows. Insets show examples of the nuclear magnetization recovery curves at two temperatures.

* Corresponding author (e43247a@nucc.cc.nagoya-u.ac.jp)

Polycrystalline samples of LaFe$_{1-y}$Co$_y$AsO$_{1-x}$F$_x$ (with $x$ always being fixed at 0.11) were prepared from initial mixtures of La, La$_2$O$_3$, LaF$_3$ and FeAs with the nominal molar ratios. Details of the preparation processes can be found in the previous papers.[4-7] The X-ray powder patterns were taken with Cu$k\alpha$ radiation. For the samples, very weak peaks of impurity phases of LaOF and LaAs were found and their molar fractions were estimated to be ~3 % and ~2.6 %, respectively. Any other impurity phases have not been detected. The superconducting diamagnetic moments were measured by a Quantum Design SQUID magnetometer with the magnetic field $H$ of 10 G under both conditions of the zero-field-cooling (ZFC) and field cooling (FC). From the data of the electrical resistivities $\rho$ and diamagnetic moments, the $T_c$ values were determined as described poreviously,[4-7] where we found that both kinds of $T_c$ values agreed well. The $^{75}$As-NMR measurements were carried out by the standard coherent pulse method, where the spectra were measured by recording the nuclear spin-echo intensity $I$ with the NMR frequency or applied magnetic field being changed stepwise. NMR longitudinal relaxation rates $1/T_1$ were measured by taking the integrated intensities of the spin echo signals against the time $t$ elapsed after saturation pulse.

Here, we add details on our samples of LaFe$_{1-y}$Co$_y$AsO$_{0.89}$F$_{0.11}$.[4-7] The superconductivity has bee n found only in the region of $y \leq y_c \sim 0.05$, and for $y > y_c$, the system is a good metal but nonsuperconducting. Even though $T_c$ vanishes with increasing $y$ at $y_c$, we can easily conclude that doped Co atoms are not acting as the pair breaking centers from the following facts. Although the linear $y$ dependence of the lattice parameter $c$ guarantees that the Co-doping is successful, $T_c$ does not have a systematic correlation with $y$. It indicates that the disappearance of the superconductivity is not due to the pair breaking effect for the superconductors with node(s) of $\Delta$, as already stated above. (If it is due to the pair breaking effect, the $T_c$ decrease should be much larger than the observation and linear in $y$.)

In Fig. 1, the NMR $l/T_1$ is plotted against $T$ in logarithmic scales for the samples with $y=0.0$ and 0.0075. The arrows indicate the $T_c$ values (23.0 and 21.5 K for the former and the latter, respectively) under the measuring magnetic fields $H$. Inset shows examples of the nuclear magnetization recovery curves taken at 6 and 30 K for the sample with $y=0.0$. The present data has following features. First, there is no coherence peak consistently with other reported data. Second, the decrease of $l/T_1$ with decreasing $T$ is rather rapid, much more rapid than those reported in refs.16-20: As summarized in Fig. 2(a), the $l/T_1$ data of many groups can roughly be described by the relation $1/T_1 \propto T^{2.5-3.0}$ in the wide $T$ region from the temperature immediately below $T_c$ down to $(0.1\sim0.2)T_c$, while the present data are described by an approximate relation $1/T_1 \propto T^6$ in the region of $T$ between ~0.4$T_c$ and $T_c$. Below 0.4$T_c$, the value of the $T$ derivative of the $1/T_1$-$T$ curve is smaller than that in region $T > 0.4T_c$. We do not know if we can express the low $T$ behavior in the form of $1/T_1 \propto T^\alpha$, because it is not easy to accurately estimate the intrinsic value of $1/T_1$ in the region of very small relaxation rate.

In Fig. 2(b), the $1/T_1$-$T$ curves of the samples of LaFe$_{1-y}$Co$_y$AsO$_{1-x}$F$_x$ with $y=0.0$ and 0.0075 are shown together with the data reported recently by Fukazawa et al.[27] for Ba$_{0.6}$K$_{0.4}$Fe$_2$As$_2$ ($T_c$ =38 K at $H$=0 and 34 K at $H$= 5.97 T). The $T$ dependences of these samples are very similar, but they are quite different from the behavior of $1/T_1 \propto T^{2.5\text{-}3.0}$ reported by many other group.[16-20] The results of the present samples and $Ba_{0.6}K_{0.4}Fe_2As_2$, shown in Figs. 2(a) and 2(b) indicate that the relation, $1/T_1 \propto T^{2.5\text{-}3.0}$ reported previously for many samples is not a common feature of Fe pnictide systems.

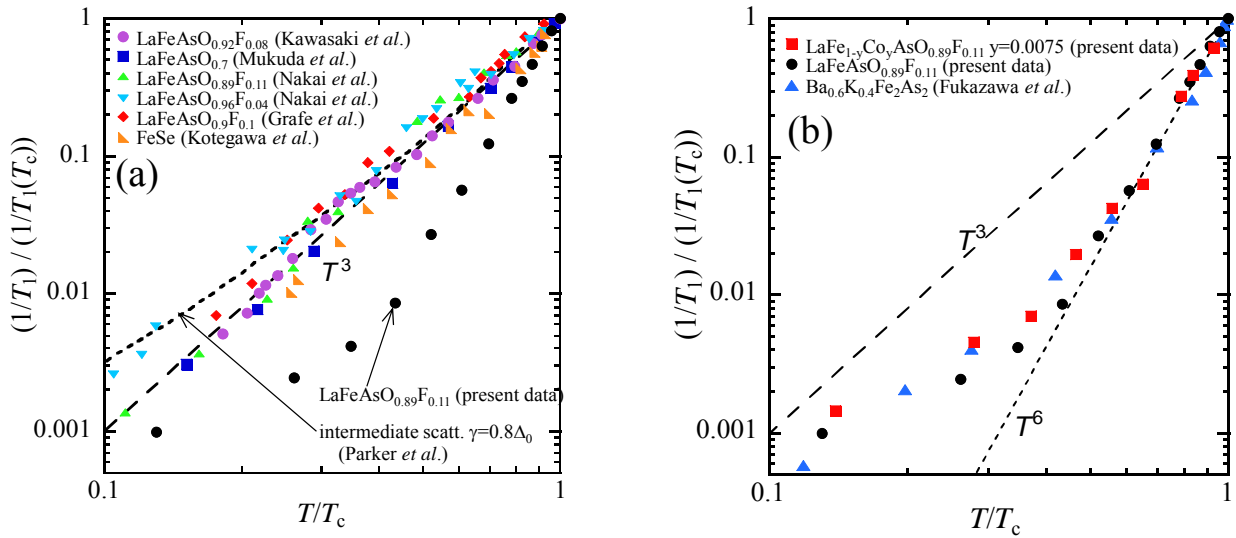

Fig. 2. (a) NMR longitudinal relaxation rate $1/T_1$ of the sample of LaFeAsO$_{1-x}$F$_x$ ($x$=0.11) is plotted against $T/T_c$ together with those reported by many groups. They are scaled by the value at $T_c$. The broken line shows the $T^3$ dependence of $1/T_1$, while the dotted curve shows the $T$ dependence of $1/T_1$ deduced by Parker et al.,[28] by considering the effect of impurities with the intermediate scattering strength. (b) $^{75}$As NMR longitudinal relaxation rates $1/T_1$ scaled by the values at $T_c$ are shown against $T/T_c$ for LaFe$_{1-y}$Co$_y$AsO$_{1-x}$F$_x$ ($y$=0.0 and 0.0075). The data reported by Fukazawa et al.[27] The $T$ dependences of these data are quite different from the $T^{2.5-3.0}$ dependence reported by many groups (see Fig. 2(a)), indicating that the $T^{2.5-3.0}$ dependence is not universal.

The relation $1/T_1 \propto T^{2.5\text{-}3.0}$ may have to be considered, as already stated above, to be due to an origin different from line nodes. Parker et al.[28] have considered effects of impurities with intermediate scattering strength to explain the relation by the $s_\pm$-symmetry, which does not have nodes of the order parameters $\Delta$ on the Fermi surfaces but has the sign difference of $\Delta$ between the disconnected Fermi surfaces around $\Gamma$ and M points in the reciprocal space. However, in their arguments, they have not considered the $T_c$-decrease caused by the scattering. As Senga and Kontani[29, 30] have shown, for impurities with intermediate scattering strength, the $T_c$ decrease is expected to be very large in contrast to the case of the unitary scattering. Because the $T^{2.5\text{-}3.0}$ behavior has been observed even for a sample of $LaFeAsO_{1-x}F_x$ with $T_c$ value as high as 28 K,[16-18, 20] the theory by Parker et al. does not seem to be applicable to the actual experimental situation. On this point, Nagai et al. have explained the $T^{2.5\text{-}3.0}$ behavior by considering the significant anisotropy of $\Delta$.[31] However, the results of ARPES study on $Ba_{0.6}K_{0.4}Fe_2As_2$[26] indicate that $\Delta$ does not have such a large anisotropy. Although we do not know whether the order parameters $\Delta$ of $LaFe_{1-y}Co_yAsO_{1-x}F_x$ have large anisotropies, it is not easy to simply attribute the origin of the $T^{2.5\text{-}3.0}$ behavior to the anisotropic order parameters of these superconductors.

Now, we know that the impurity-doping effect is not compatible with the order parameters $\Delta$ having node(s) on the Fermi surface and that it is not compatible, either, with nodeless order parameters $\Delta$ on the Fermi surfaces but with their sign difference between disconnected Fermi surfaces.[5-7] We have shown that the $T^{2.5\text{-}3.0}$ dependence of $1/T_1$ cannot be considered to be the experimental evidence for the $s_\pm$-symmetry of the order parameter. Then, what is the possible symmetry of the order parameter? One way to explain the above results are, although it is not easy, to attribute the $T^{2.5\text{-}3.0}$ behavior to certain extrinsic effects. In that case, the impurity effect on $T_c$ and very steep $T$ dependence ($T^6$ dependence) of $1/T_1$ can roughly be understood by introducing the unitary impurity-scattering.[28-30]

If we introduce the simple s-wave symmetry, we have to explain the absence of the coherence peak, which may not be very difficult, because the energy broadening of the $3d$ electrons at around $T_c$ is large enough to wipe out the peak.[32] It may be also supported by the significant increase of the thermal conductivity induced by the occurrence of the superconductivity.[33] Then, the very fast decrease of $1/T_1$ observed here with decreasing $T$ is considered to originate from the s-wave symmetry of the order parameter. However, at this point, we do not know how the relation $1/T_1 \propto T^{2.5\text{-}3.0}$ reported by several groups can be understood by the s-wave symmetry. In this sense, experiments sensitive to the signs of the order parameters are desirable. On this point, Maier and Scalapino[34] and Korshunov and Eremin[35] discussed the possible existence of the resonance peak in the spectral function of the magnetic excitation for the case of the $s_\pm$-symmetry of the order parameter. Experimental data of the spectral function[22-24] should be carefully examined by considering various other effects induced by the occurrence of the superconductivity.

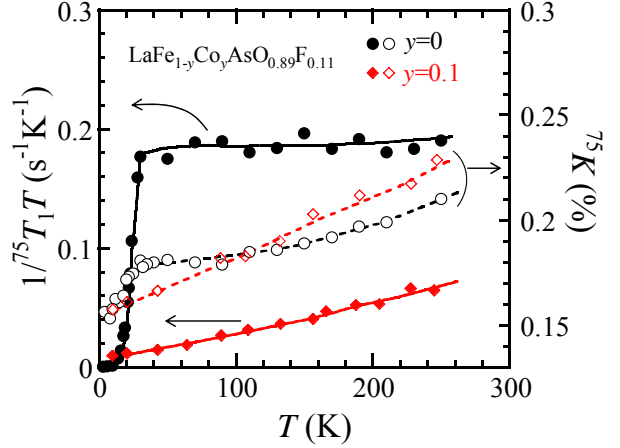

Fig. 3  NMR longitudinal relaxation rates $1/T_1T$ divided by $T$ and Knight shifts $K$ are shown against $T$ for the samples of $LaFe_{1-y}Co_yAsO_{1-x}F_x$ with $y=0.0$ and $0.1$.

In Fig. 3, NMR longitudinal relaxation rates $1/T_1T$ and Knight shifts $K$ are shown for the samples of $LaFe_{1-y}Co_yAsO_{1-x}F_x$ with $y=0.0$ and $0.1$. From the figure, we can find following characteristics. First, the $T$ dependences of these two kinds of quantities of each sample are not completely identical, but they are so similar as to exclude a possibility that there exists a sharp peak of the magnetic susceptibility in the reciprocal space. It indicates that the electron system is not very close to the magnetic instability. Second, $1/T_1T$ observed for $y=0.1$ is almost in proportion to $T$ (precisely speaking, $1/T_1T$-$T$ curve is slightly concave in the low-$T$ region and crosses the vertical axis at a small but finite value.). It seems to be understood by considering the change of the electronic density of states $N(\varepsilon)$ near the Fermi energy $\varepsilon_F$.[36, 37] Both samples have the Fermi energy near the dip of the density of states $N(\varepsilon)$, where $\varepsilon_F$ for $y=0.1$ is closer to the dip energy. However, it is interesting to examine if the $T$ dependence of these quantities can be explained by the change of $N(\varepsilon)$.

In summary, we have shown that there are experimental data of $1/T_1T$ with $T^6$ dependence in the region of $T$ between ~$0.4T_c$ and $T_c$, indicating that the $T^{2.5\text{-}3.0}$ dependence is not universal for Fe pnictide superconductors. Based on this and other related results, arguments on the superconducting order parameter have been presented.

**Acknowledgments**

The authors thank Prof. H. Kontani for fruitful discussion. The work is supported by Grants-in-Aid for Scientific Research from the Japan Society for the Pro motion of Science (JSPS), Grants-in-Aid on Priority Area from the Ministry of Education, Culture, Sports, Science and Technology and JST, TRIP.